\documentstyle[12pt,a4wide]{article}
\begin{document}
\begin{center}
{\Large Electron-spin-resonance in the doped spin-Peierls compound
Cu$_{1-x}$Ni$_{x}$GeO$_3$ }

\vspace{10mm}
V.~N.~Glazkov, A.~I.~Smirnov

{\it P.L.Kapitza Institute for Physical Problems RAS, 117334
Moscow,Russia}

\vspace{5mm}
O.~A.~Petrenko, D.~McK.~Paul

{\it Department of Physics, University of Warwick, Coventry CV4~7AL,
 UK}

\vspace{5mm}
A.~G.~Vetkin

{\it M.V.Lomonosov Moscow State University, 119899 Moscow, Russia }

\vspace{5mm}
R.~M.~Eremina

{\it E.K.Zavoisky Physical Technical Institute, 420029 Kazan, Russia}
\end{center}

\vspace*{2mm}

{\it Short title}:
ESR in doped spin-Peierls compound Cu$_{1-x}$Ni$_{x}$GeO$_3$
\vspace*{2mm}

\vspace*{20mm}

{\bf Abstract} \\
{\small
Electron spin resonance studies of the Ni-doped spin-Peierls compound
CuGeO$_3$  has been  performed in the  frequency  range  9-75~GHz and
temperature interval 1.3-20~K.  An anomalous  temperature  dependence
of the $g$-factor below the spin-Peierls temperature was observed for
doped  samples.  At low  temperatures  the $g$-factor is much smaller
than the value expected for Cu$^{2+}$ and Ni$^{2+}$ ions and is much
more  anisotropic  than  for an  undoped  crystal.  This  anomaly  is
explained by the formation of magnetic  clusters around the Ni$^{2+}$
ions within a  nonmagnetic  spin-Peierls  matrix.  The  formation  of
magnetic  clusters is  confirmed  by the  observation  of a nonlinear
static magnetic susceptibility at low temperatures.

The reduction of the spin-Peierls transition temperature was found to
be linear in the dopant  concentration  $x$ in the range $0\leq x\leq
3.2$\%.  The transition into the antiferromagnetically ordered state,
detected earlier by neutron scattering for $x\geq 1.7$\%, was studied
by means of ESR.  For  $x$=3.2\%  a gap in the  magnetic  resonance
spectrum  is  found  below  the  Neel  temperature  and the spectrum
is  well  described  by  the  theory  of antiferromagnetic resonance
based on the molecular field approximation.  For $x$=1.7\% the
spectrum below the Neel point remained gapless.  The   gapless
spectrum of the antiferromagnetic state in weakly doped samples is
 attributed to the small value of the Neel order  parameter  and  to
the magnetically disordered  spin-Peierls background.}

\vspace{10mm}

\newpage

\section{INTRODUCTION}

The  magnetic  properties  of  crystals  of the  quasi-onedimensional
magnet  CuGeO$_3$  have been  extensively  studied since Hase {\it et
al}~\cite{Hase}  reported that this  compound is the first  inorganic
spin-Peierls  material.  The spin-Peierls  transition  occurs because
the $S=1/2$  Heisenberg  antiferromagnetic  chains are unstable  when
coupled to a three-dimensional phonon  field~\cite{Pytte}.  Below the
transition  temperature  the magnetic  chains are  dimerised  and the
distance  between  the  neighbouring  magnetic  ions  as well  as the
exchange integral alternate.  The positions of pairs of magnetic ions
after the dimerisation are correlated  between  neighbouring  chains.
Thus the dimers construct an ordered sublattice.

The  dimerization  of the magnetic ions in chains  arranged along the
$c$-direction of the  orthorhombic  crystal leads to the formation of
the  non-magnetic  ground state  separated  from the excited  triplet
states   by   an   energy    gap    $\Delta\approx    2$~meV~$\approx
23$~K~\cite{Nishi}.  The unit cell in the dimerised  state is doubled
along $a$ and $c$  directions and the  intrachain  exchange  integral
takes   the   alternating   values   $J_{1,2}=(1\pm\delta)J_c$.  Here
$J_c$=10.2~meV    is   the   intrachain    exchange    integral   and
$\delta\approx0.04$  is a distortion  parameter~\cite{Regnault}.  The
existence of such  dimerization  was confirmed by the  observation of
additional      reflections      by      X-ray~\cite{Hirota}      and
neutron~\cite{Regnault}  diffraction.  Since  the  thermally  excited
triplet states are separated from a non-magnetic ground state by an
energy gap, the number of excitations and the magnetic susceptibility
should  decrease and tend to zero below the  transition  temperature.
Both    static    susceptibility     measurements~\cite{Hase}     and
ESR-studies~\cite{Oseroff,Honda,Smirnov}  performed  on  crystals  of
pure CuGeO$_3$ showed that the susceptibility rapidly decreases below
$T_{SP}=$14.5~K.  The transition  temperature to the dimerised
state was obtained from the initial decrease in susceptibility.

CuGeO$_3$  is not a perfect one  dimensional  magnet, the  intrachain
exchange  $J_c$ is larger but not much  stronger than the  interchain
exchanges    $J_b=0.1J_c$    and    $J_a=-0.01J_c$~\cite{Nishi}.    A
next-nearest-neighbour  exchange  along the chains with a significant
value  of  $J'  \approx   0.36J_c$  also  probably   exists  in  this
compound~\cite{Riera,Braden,Khomskii}.  If  the  crystal  lattice  of
CuGeO$_3$ was harder, this crystal would be antiferromagnetic  due to
the  intra-  and  interchain   exchange   interactions  with  a  Neel
temperature of about
$T_N\sim(J_cJ_b)^{\frac{1}{2}}\sim10$~K~\cite{Hennessy}.   But    the
spin-Peierls    state   wins   the   competition    with   long-range
antiferromagnetic  order.  Doping  by  impurities  makes  long  range
antiferromagnetic  order possible again.  The presence of impurities
diminishes  the  temperature of the  spin-Peierls  transition  and at
lower  temperatures a Neel ordering  occurs~\cite{Regnault1,Lussier}.
The spin-Peierls  dimerization and the antiferromagnetism  were found
to coexist  in doped  crystals.  This  phenomenon  was  explained  by
Fukuyama   {\it  et   al}~\cite{Fukuyama}   and   Khomskii   {\it  et
al}~\cite{Khomskii},   who  showed  by  different  methods  that  the
impurity or the defect in the dimerization  should be surrounded by a
region of  antiferromagnetically  correlated  spins, forming magnetic
clusters,  or solitons.  The tails of these  solitons  could  overlap
producing  long range magnetic  order.  From such a point of view the
reduction of the  spin-Peierls  temperature  is  proportional  to the
concentration of the dopant for small concentrations.

The  impurity  or  the  point  magnetic   defect  inserted  into  the
nonmagnetic  spin-Peierls  matrix  generates a ``many  spin''  object
consisting of the impurity itself and of several neighbouring Cu-ions
spins.  The  magnetic  object  formed  due  to  the  presence  of the
impurity is therefore a mesoscopic type object containing a number of
spins which is  intermediate  between a microscopic  and  macroscopic
systems, while the magnetisation remains microscopic.

Single   crystals  of  CuGeO$_3$  with  the  following   substituting
impurities    were    investigated:   Si,   Ti   on    Ge-sites~\cite
{Regnault1,Weiden},         Zn,          Mg~\cite{Hase1,Hase2,Hase3},
Ni~\cite{Petrenko,Koide} and Co~\cite{Anderson} on the Cu-sites.  The
suppression  of  the   spin-Peierls   transition   and  occurrence  of
antiferromagnetic  order at low  temperatures are common features for
different  dopants.  In  the  antiferromagnetic  phase  the  magnetic
moment per Cu-ion is strongly  reduced.  This reduction  depends upon
the type of dopant and their  concentration.  For the 3.2 \% Ni-doped
crystal the effective moment is $\mu_{eff}=0.16\pm0.03\mu_B$  while it
decreases   to    $0.06\pm0.03\mu_B$    for   the   1.7\%    Ni-doped
sample~\cite{Petrenko}.    For    the    3.2\%    Zn-doped    crystal
$\mu_{eff}\approx0.2\mu_B$~\cite{Sasago}.  The  direction of the easy
axis of the  antiferromagnetically  ordered state also depends on the
dopant:  the  easy  axis  is   directed   along  the   $c$-axis   for
Zn~\cite{Smirnov}, Si~\cite{Nojiri}, Co~\cite{Anderson} while for the
Ni-doped crystals it is directed along the $a$-axis~\cite{Koide}.

The aim of the present  paper was the study of the ESR  signals  from
the  impurity  seeded  magnetic  clusters  and  the  search  for  the
antiferromagnetic   resonance  (AFMR)  in  Ni-doped  single  crystals
Cu$_{1-x}$Ni$_x$GeO$_3$.  These  were the same  crystals  which  were
investigated earlier by neutron scattering~\cite{Petrenko}.

\section{EXPERIMENTAL DETAILS AND SAMPLES}

In  our  experiments  we  used  a  set  of  ESR  spectrometers   with
transmission  type  cavities in the  frequency  range  8-80~GHz.  The
microwave  cavities  were placed in a  hermetically  isolated  volume
immersed  in a liquid  helium bath and filled with a small  amount of
He-gas, enabling one to vary the temperature of the cavity containing
the  sample  in  the  range  1.3 -  20~K  .  The  magnetic  resonance
absorption  lines  were  recorded   through  the  dependence  of  the
transmitted  microwave  power  on the  applied  magnetic  field.  The
reduction  of the  transmitted  power is  proportional  to the  power
absorbed  by the  sample  when  the  absorbed  power  is low.  In the
paramagnetic  state the intensity of the absorption  integrated  over
magnetic field is  proportional to the static  susceptibility  of the
spin system.

The  magnetization  curves were obtained using an Oxford  Instruments
vibrating sample magnetometer.

Single  crystals  of   Cu$_{1-x}$Ni$_x$GeO$_3$  with  $x$=0.017   and
$x$=0.032 were  produced by  the crystal  growth procedure  described
in~\cite{Petrenko}.   These were  the same  samples used  for neutron
scattering experiments.  These  experiments showed the appearance  of
antiferromagnetic  order  at  $  T_N$=2.3~K  in  the  1.7\%  Ni-doped
crystal  and  at  4.2~K  for  $x$=  3.2\%  .  The  transition  to the
spin-Peierls  state  at  $T_{SP}=11.5$~K  was  observed for the 1.7\%
Ni-doped  crystal,  while  for  the  sample  containing  3.2\%Ni  the
spin-Peierls transition was not clearly observed.  The values of  the
exchange  integrals  $J_b=0.7\pm0.1$~meV,  $J_c=1.8\pm0.3$~meV and of
the spin wave energy gap $\Delta\approx$ 0.18~meV were obtained  from
the  dispersion  curves  of  the  magnetic  excitations  in the 3.2\%
Ni-doped CuGeO$_3$  at $T=$1.5~K.   To investigate  the influence  of
the Ni-doping  at low  concentrations, when  the dopant  atoms do not
interact,  crystals  of  Cu$_{1-x}$Ni$_x$GeO$_3$  with  $x\leq 0.005$
were grown  by recrystallization  of ceramic  samples in  air using a
horizontal floating-zone method.  The growth rate was 5-7~mm/h.   The
ceramic   samples   were   prepared   by   annealing   in   air    of
nonstoichiometric mixture of CuO, GeO$_2$ and Ni$_2$O$_3$ at  950$^o$
C for 24 hours.

\section{EXPERIMENTAL RESULTS }

\subsection{The temperature evolution of the ESR line and the AFMR
spectrum}

The evolution of the ESR line with temperature for the pure crystals of
CuGeO$_3$ is described in~\cite{Oseroff,Honda,Smirnov}.  The intensity
of the resonance absorption diminishes rapidly below $T_{SP}$.  The
nonmonotonic change of $g$-factors $g_a$, $g_b$ and $g_c$  takes place
at the temperature variation in the range between 14.5 and 4~K.  This
change occurs near the values which are close to $g$=2.1 and the magnitude
of this change does not exceed 4~\% for $g_a$ and $g_b$ and 1~percent
for $g_c$.  The additional line with $g_a=1.82$, $g_c=$1.45, $g_b=$1.86
occurs below 4~K.  The main line splits into three narrow lines at low
temperatures.  At a temperature of 4~K the ESR integral intensity of
pure samples from our set of crystals is about 3$\times$10$^{-3}$ of
the intensity of a paramagnet with one electron spin per Cu-ion.  The
residual ESR signal of the pure crystals is attributed to defects in
the structure and to residual impurity content, as well as to the
boundaries of the spin-Peierls domains~\cite{Smirnov1}.

The intensity of the ESR signals  observed at low temperatures in  the
Ni-doped  samples  is larger  than the  intensity  of the ESR in  pure
samples.  The value of the  intensity  for the $x$=0.5\%   corresponds
approximately to the concentration of the inserted  impurities,  while
for $x  \geq$1.7\%  the  intensity is less than that of a   paramagnet
with the  corresponding  amount of $S=$1 spins.  The evolution of  the
ESR line with  temperature for samples with different   concentrations
of impurity is shown in Figures~1, 2, 3.

In  contrast  to the  pure  material,  the  value  of the  $g$-factor
strongly  depends  on the  temperature  below  $T_{SP}$  in  Ni-doped
crystals.  The resonance  field is shifted  toward higher fields with
decreasing temperature and $g$-factor diminishes by about 20\%.  For
the smallest concentration,  $x$=0.005, (Figure~1) an additional weak
line  with a  $g$-factor  of  about  2  arises  at low  temperatures,
indicating an ESR spectrum which is intermediate between the pure and
doped samples.

The sample doped with 1.7\% Ni showed an analogous  temperature shift
of  the  ESR  line  (Figure~2).  The  linewidth  has a  maximum  at a
temperature of about 2.5~K (see Figure~4), which is close to the Neel
temperature  for this  crystal.  There is no observable  shift of the
line  position  between  the Neel  temperature  2.3~K and the  lowest
temperature  of  our   experiments   1.3~K  in  the  frequency  range
18-60~GHz.

The sample containing 3.2\%  Ni demonstrated a transformation  from a
single-line spectrum  into the  spectrum typical  of an  orthorhombic
antiferromagnet with several resonant  lines.  Three lines  appear at
this  transformation  when  the  magnetic  field  is  parallel to the
direction of spin ordering  (Figure~3), and a gap opens  in
the absorption spectrum for  other directions of the  magnetic field.
This  transformation  occurs  at   4~K  corresponding  to  the   Neel
temperature  for  this  sample.   Above  the  Neel  temperature,  the
resonance field slightly  increases as the temperature  decreases, but
this shift is smaller than for more lightly doped samples. The temperature
evolution of the resonance field is given in Figure~5.

We  performed measurements  of  the  resonance  field  in   the
frequency   range   18-75~GHz.     The   resonance    frequency-field
dependencies for 0.5\% Ni and 1.7\% Ni-doped samples were linear  and
gapless in the temperature interval  between 1.3 and 20~K. A  typical
dependence of the resonance frequency  $f$ on the magnetic field  $H$
is  given  in   Figure~6.   The   temperature  dependencies  of   the
$g$-factors taken  at different  frequencies are  given in  Figure~7.
The sample  doped with  3.2\%~Ni is  presented here  only at  $T \geq
T_N$, where the spectrum is  gapless.  The values of  $g$-factors are
obtained by the relation:

$g=2 f/\gamma H$ with $\gamma=28$ GHz/T.

Note that  the   deviation   of the   $g$-factors   with  temperature
does not  tend towards  the value  of the  $g$-factor   for Ni$^{2+}$
ions which has the  value about 2.3.   At $T>T_{SP}$ there is  also a
remarkable  deviation of  $g$-factor  for the  doped  crystals   from
the   value  of  the  pure  material.   This deviation increases with
increasing concentration.

The AFMR spectrum of the 3.2\% Ni doped sample at $T$=1.8~K which has
well pronounced gaps of 22~GHz and 33~GHz is presented in Figure~8.

\subsection{Spin-Peierls transition}

The  drop of the  integrated  intensity  of the ESR  line  marks  the
temperature of the  spin-Peierls  transition  both for pure and doped
crystals.  The appropriate  temperature  dependence of the integrated
intensity  is shown in  Figures~9-a,b.  The  transition  temperatures
obtained from the ESR data are given in the Table~1.

\vspace{5mm}

{\bf Table 1} {\small The temperatures of the spin-Peierls transition
obtained from the ESR intensity vs temperature dependencies for
the samples of Cu$_{(1-x)}$Ni$_x$GeO$_3$}
\begin{center}
\begin{tabular}{|c|c|c|c|c|}
\hline
$x$         & 0           & 0.5\%        & 1.7\%        & 3.2\%     \\
\hline
$T_{SP}$,~K &14.0$\pm$0.5 & 12.5$\pm$0.5 & 11.5$\pm$0.5 & 8.0$\pm$1 \\
\hline
\end{tabular}
\end{center}

The values of $T_{SP}$ for doped  crystals are smaller than those for
pure  CuGeO$_3$.  The value of  $T_{SP}$  for the 1.7\% doped  sample
agrees well with the neutron scattering  data~\cite{Petrenko},  while
for the 3.2\% doped sample the neutron scattering data cannot be used
for the  determination  of  $T_{SP}$  because of the almost  complete
collapse of the spin-Peierls excitations.

\subsection{Magnetization curves}

The  magnetization  {\it vs}  magnetic  field  was  measured  at
different temperatures  for the  1.7\%  Ni  doped  sample  to
investigate  the magnetic  properties  of the sample in the
temperature  range of the strong   $g$-factor   evolution.
Figure~10   illustrates   the  M(H) dependencies at different
temperatures for $H \parallel a$.  For the clear  demonstration  of
the nonlinear  contribution a fixed linear part is subtracted  and the
difference  $\Delta M$ is  plotted.  Significant nonlinearity arises
at low (T$<$6~K) temperatures.  The inset of this Figure
demonstrates  the  spin-flop   transition  at  $H$=0.3~T,  $H
\parallel  a$.  For $H  \parallel b$ and $H  \parallel  c$ the $M(H)$
curves  are  similar  except  for  the  spin-flop  steps,  which  are
observable only for $H \parallel a$.  The nonlinear parts for the $a$
and  $c$-directions  of the  magnetic  field  are the  same,  for the
$b$-direction the nonlinear contribution is about 50\% greater.

\section{DISCUSSION}

\subsection{x-T phase diagram}

The values of the  spin-Peierls  transition  temperature and the Neel
points  obtained from ESR intensity are in a good agreement  with the
$x-T$   phase   diagram   for   the   Ni-doped   samples    presented
in~\cite{diagramma}  with a linear  decrease of $T_{SP}$  with $x$ in
the interval 1.7-2.9\%.  Our additional points at $x=$0.5\% and 3.2\%
extend the experimental verification of the linear region.

\subsection{Antiferromagnetic resonance}

Doped   crystals,   which   demonstrated   the  Neel   order  at  low
temperatures,  show different  types of magnetic  resonance  spectra.
The sample  doped  with 1.7\% Ni has the Neel point at 2.3~K, but did
not show a  characteristic  gap in the  magnetic  resonance  spectrum
(Figure~6).  The  sample  doped  with  3.2\%  Ni  has  a gap  in
the resonance spectrum (Figure~8).  This spectrum may be described
as the AFMR spectrum of an orthorhombic antiferromagnet,
obtained by means of the Landau-Lifshitz  equations of sublattice
magnetizations motion  within a  molecular  field  approximation.
Taking  the  axes $a,c,b$ as easy,  second easy and the hard axis
correspondingly,  we obtain the resonance frequencies $\nu_{1,2}$ as
follows~\cite{formula}:

\begin{eqnarray}
\lefteqn{H\parallel a , H<H_{SF}:}\nonumber\\
\lefteqn{(\nu_{1,2}/\gamma)^2=\frac{1}{2}[(1+\alpha^2)H^2+C_1+C_2\pm}\nonumber\\
& &
((1-\alpha^2)^2H^4+2(1+\alpha)^2(C_1+C_2)H^2+(C_1-C_2)^2)^{\frac{1}{2}}]\\
\lefteqn{H\parallel a , H>H_{SF}:}\nonumber\\ & &
(\nu_1/\gamma)^2=H^2-C_1\\ & & (\nu_2/\gamma)^2 =C_2-C_1\nonumber\\
\lefteqn{H\parallel c :}\nonumber\\ & & (\nu_1/\gamma)^2=H^2+C_1\\
& &
(\nu_2/\gamma)^2= C_2\nonumber\\ \lefteqn{H\parallel b :}\nonumber\\
& & (\nu_1/\gamma)^2=H^2+C_2\\ & & (\nu_2/\gamma)^2= C_1\nonumber
\end{eqnarray}
\\
here  $C_{1,2}=2H_eH_{a1,a2}$  ($H_e$,$H_{a1,a2}$  are exchange field
and fields of anisotropy respectively), $H_{SF}=(2H_eH_{a1}/\alpha)^{1/2}$
is the spin-flop field and  $\alpha=1-\chi_{\parallel}/\chi_{\perp}$.
The  applied  magnetic  field  $H$  and  the  anisotropy  fields  are
considered to be much smaller than the exchange field.

By fitting our data  according  to  equations  (1-4) we obtained  the
following  parameters of the AFMR spectrum:  $H_{SF}=$1.2$\pm$0.05~T,
$\gamma=$      24.6~GHz/T,      $2H_eH_{a1}=0.85\pm      0.10$~T$^2$,
$2H_eH_{a2}=1.80$~T$^2$,  $\alpha=$0.75$\pm$0.2.  The gap  values for
the  second-easy  and hard axis  directions of the magnetic field are
22~GHz and 33~GHz (0.11~meV and 0.17~meV)  respectively, which are in
agreement   with  the  value  of  0.18~meV   obtained   from  neutron
scattering~\cite{Petrenko}.

The spin-flop  magnetic field  is characterized  by the  wide band of
absorption instead of the single resonance frequency (see  Figure~8).
According  to  our  data  $H_{SF}$=1.2~T.  This  value  is  in   good
agreement  with  the  value  of  the  spin-flop  field  1.1~T   found
in~\cite{Koide}  from  the  magnetization  curves  for  the  Ni-doped
crystal with x=0.033.

From the exchange integral values $J_b\approx0.7$~meV,
$J_c\approx1.8$~meV described in Section 2 and assuming $S=1/2$ and
$H_e=4(J_c+J_b)S/(g\mu_B)$ we estimate the exchange field value $H_e
\approx $ 27~T.  Further, from our data on the AFMR gap we deduce the
value of the anisotropy fields:  $H_{a1}\approx0.017$~T and
$H_{a2}\approx0.036$~T.

The single line ESR spectrum transforms into the AFMR spectrum with a
gap for $x=$3.2\% but for $x=$1.7\% the  transition to the Neel state
is marked only by the maximum of the linewidth.  If the  conventional
approach to the AFMR frequency  derivation  were valid for the sample
containing  1.7\% Ni, the observed value of the spin flop  transition
(Figure~10)  would  correspond  to an AFMR  gap of  7.5~GHz.  For the
frequency  18~GHz at $H  \parallel  a$ it should shift the  resonance
field by 0.05~T  towards  higher  fields and for $H  \parallel  c$ to
lower  fields.  This shift  should be visible in our  experiments  by
lowering  temperature from $T_N=$2.3 to 1.3~K.  Nevertheless no shift
exceeding 0.02~T was observed.  This discrepancy between the
static   and    dynamic    properties    of   the    impurity-induced
antiferromagnetic   state  within  the  spin-Peierls  matrix  may  be
attributed  to the low  value  of the  order  parameter  for the Neel
state.  The sublattice  magnetization equals only 0.06 of the nominal
value. Usual procedure for deriving the AFMR frequency
is therefore not valid because of the  assumption  that the spins are
arranged  in the form of hard  sublattices,  without a  reservoir  of
disordered spins.  In the case of the coexistence of the spin-Peierls
and  Neel  ordered  states  the  magnetically  ordered  part  of  the
magnetization  should  interact  with  the  disordered   spin-Peierls
background  via  the  strong  exchange  interaction  $J_c$.  It  will
probably cause an unusual type of magnetic resonance  frequency or an
overdamped mode.

\subsection{Spin clusters and magnetic resonance}

The value  of $g$-factor  in Ni-doped  samples differs  strongly from
the corresponding values of pure crystals and of crystals doped  with
other impurities~\cite{Nojiri,Hase3}.   This difference  takes  place
both above  and below  $T_{SP}$.   Below the  spin-Peierls transition
there is also a strong anisotropy in the $g$-factor.  Because  of
the pronounced  difference between  the observed  $g$-factor and that
expected for individual  Ni- and Cu-ions,  we consider that  clusters
of several spins coupled by the exchange interaction are  responsible
for the discrepancy.  The  $g$-factor of the cluster of  ions coupled
by  the  symmetric  Heisenberg  exchange  takes  the  averaged  value
between  the  $g$-factors  of  isolated  ions~\cite{Abragam}.   The
observed  $g$-factor  value  is well outside of this interval.  The
possible reason for  this striking  $g$-factor deviation  is the
formation of clusters  containing  several  spins  coupled  both  by
symmetric and antisymmetric exchange interactions~\cite{Belinskii}.

The formation  of clusters  should occur  in doped CuGeO$_3$-crystals
around the dopant  ions, as described  in Sec.~1, because  the defect
is  surrounded  by  several  antiferromagnetically  correlated spins.
The characteristic length of  the reduction of the  correlated spin
component on  moving  away  from  the impurity
should be about 7 interionic distances~\cite{Khomskii}.

Clusters  containing  three  $S=1/2$  ions  are  known  to display an
effective $g$-factor which  is  smaller  than  the  $g$-factor  of an
isolated ion.   The anisotropy of  the $g$-factor for  the cluster is
larger        than        that         of        the         isolated
ions~\cite{Yablokov1,Yablokov2,Belinskii}.    This  change   in   the
$g$-factor  is  described  by  taking  into  account the spin-orbital
interaction combined with  the Heisenberg exchange  in the form  of a
Dzyaloshinski-Moriya  antisymmetric  exchange  which  is allowed when
the symmetry  of the  pairs of  the interacting  ions is  low enough.
According  to~\cite{Belinskii}   the  reduction   of  the   effective
$g$-factor  is  of  the  order  of  $D/\delta  J$.   Here  D  is  the
antisymmetric exchange coefficient and  $\delta J$ is the  difference
between  the  exchange  integrals  within  a triangular spin cluster.
The presence of the  Dzyaloshinski-Moriya term in pure  CuGeO$_3$ was
proposed  in~\cite{Yamada}  to  explain  the  ESR  linewidth  at high
temperatures.     In addition,   the   Dzyaloshinski-Moriya   term
describing  the  interaction  of  two  neighbouring magnetic ions may
arise below T$_{SP}$  because of the  lowering of the  local symmetry
resulting from the dimerization.

Figure~11 shows an  impurity atom embedded  in the dimerised  matrix.
The  dimerization  is  disturbed  in  the  vicinity  of the impurity.
There are  no symmetry  centers for  Ni-Cu pairs.   The region around
the impurity atom also does not  have the symmetry center.   Therefore
Dzyaloshinski-Moriya interactions coefficients should have  different
nonzero  values  for  each  pair  of  ions  in  the  vicinity  of the
impurity.

We  consider  the  models  of  the  spin  cluster  with the final and
relative small  numbers of  spins to  describe qualitatively  the ESR
spectra    and    to    evaluate    the    coefficients    of     the
Dzyaloshinski-Moriya interaction.  The larger the number of  spins
under consideration the more realistic is the model because the
real cluster is formed on the basis of the impurity in the infinite
chain.  For the large number of spins in the model the correlated
component of the spins lying far from the Ni-spin should be strongly
reduced due to the dimerization. Therefore the magnetic properties
of the model  cluster should be  independent of the number  of
spins.  The possible  values of the total spin of the cluster
arising due to the substitution of one Cu-ion per  Ni-ion are $S$=1
and $S$=1/2.  In order to obtain $S=$1 the spin of the  Ni-ion must be
uncompensated and therefore the  Cu-ions are divided into dimers
in another way when compared to the  undisturbed chain. The
correlation of the dimerization in the neighbouring chains will be
violated in this case. For the case of the total spin $S=$1/2 the
 spin of the Cu-Ni pair is uncompensated.  This total spin value
does not cause the rearrangement of the dimers lattice  except for
the single pair Cu-Cu replaced by Cu-Ni pair.  Therefore the total
spin $S=$1/2 of the cluster corresponds to the lower energy of the
perturbation of the  dimerised lattice and the $S=$1/2 models
with the finite number of spins are probably more realistic then the
models with $S=1$.

We have analyzed  the five-  and six-spin-models for the cluster. The
five-spin-model has $S=$1 in the ground state and  the six spin
model  has $S=$1/2.  The three-spin cluster  would be the simplest
case,  but it has $S=0$  in the  ground state  and is  therefore
nonmagnetic.   The four spin model  could not include  dimers from
both sides of the Ni-ion.

The five- and six-spin-clusters are shown schematically in the
Figure~11 and are described by the Hamiltonian

\begin{equation}
{\cal H}= \sum_{i,j}^{K}{}^{'}J_{i,j}{\bf S_iS_j}+
\sum_{i,j}^{K}{}^{'} {\bf D}_{i,j}{\bf S_i \times S_j}+
\sum_{i=1, \alpha=a,b,c}^{i=K} g_{i\alpha}\mu_B H_{\alpha}
S_{i\alpha} \label{Ham}
\end{equation}
\\
Here $g_{i  \alpha}$  are  $g$-tensor  components  of Cu and
Ni-ions, $\mu_B$ is the Bohr magneton.  $J_{ij}$ are the
nearest-neighbour and the next-nearest-neighbour  exchange integrals
and ${\bf D}_{ij}$ are the vectors of the antisymmetric  exchange.
The different pairs of ions are taken only once in the sums
$\sum {}^{'}$.  $K$ is the number of spins equal to 5 or 6.

In the following analysis we shall assume the ${\bf D}_{ij}$ vectors to
be  parallel to each other and perpendicular to $c$ axes of the
crystal.  For the five spin cluster the ions are labeled
Cu$_5$-Cu$_4$-Ni$_1$-Cu$_2$-Cu$_3$ and the nonzero $J_{ij}$ are taken
as follows. $J_{54}=J_{23}$=10.6~meV are the Cu-Cu exchange integrals;
$J_{12}=J_{14}=$5~meV are the exchange integrals between Ni and Cu
ions, they are estimated as a half of the Cu-Cu exchange
integral~\cite{Eremin}; $J_{42}$=3.6~meV is the next-nearest-neighbour
exchange in the pure material~\cite{Riera}); $J_{13}$ and $J_{15}$, the
next-nearest-neighbour exchange of the Ni-ion are taken to be 2~meV.
The ${\bf D}_{ij}$ vectors are supposed to be parallel to each other
and perpendicular to the chains.

The diagonalization of the  Hamiltonian~(\ref{Ham})  gives the
energy levels for S=1 states of the five-spin-cluster depending on
the $D_{ij}$ values and magnetic field.  The levels characterized
with $S_z=0,\pm 1$ are spilt by the $D_{ij}$ terms of (\ref{Ham}). The
transitions with the momentum change  of $\pm \hbar$ are therefore
separated by a gap of 1.0 meV in zero field. The formulae determining
the gap and the values of $g$-factor are given in the Appendix. This
gap is the main feature differentiating the $S$=1 and $S$=1/2 cases of five-
and six-spin models.

The lowest $S=$1/2 states of the of the six-ion-clusters are also
separated by a gap which depends on the exchange integrals and $D_{ij}$
values. A magnetic field splits these levels, this splitting depends on
magnetic field and $D_{ij}$. However, the transitions between the
sublevels with  the  $\pm \hbar$ momentum change remain gapless.  The
modification of the  ESR spectrum at low frequencies is restricted here
to the renormalization of the effective $g$-factors.  The energy levels
are obtained by the diagonalization of the energy matrix 10$\times$10
for the $S=$1/2 states of the six-spin-cluster.  For the cluster with
 the ions labelled as Cu$_5$-Cu$_6$-Ni$_1$-Cu$_2$-Cu$_3$-Cu$_4$ with
the following values of the exchange integrals $J_{12}$=5~meV,
$J_{16}$=5~meV, $J_{23}$=9.8~meV, $J_{34}= J_{56}$=10.6~meV
the terms dominating in g-factors at magnetic field perpendicular and
parallel to Dzyaloshinski-Moriya vectors are:

\begin{equation}
g_{\perp }\approx \frac{4g_{Ni}-g_{Cu}}3-\frac{4g_{Ni}+5g_{Cu}}9\frac{%
(4D_{16}+3D_{65})^2}{54E_{21}^2}
\end{equation}

\begin{equation}
g_{\parallel} \approx \frac{4g_{Ni}-g_{Cu}}{3}
\end{equation}

where

\begin{equation}
E_{21} \approx J_{56} -\frac{2}{3}J_{61}+\frac{1}{6}J_{62}
-\frac{5}{9}J_{15}
\end{equation}
 is the energy interval between the ground state and the lower
 excited state of the cluster.

To obtain $g_{\perp }$=1.6 as observed in our experiments we have to assume
$D_{16}\approx D_{65}\approx $3meV. The exact diagonalization of the matrix
gives the close values of $|D_{ij}|=$ 3.4meV.

The values of the averaged spin projection at the site for six-spin-cluster
obtained by the described procedure are given in the Table 2 and are shown
schematically in Figure~11. The following numerical values are used at this
procedure in addition to the exchange integrals given above: D$_{12}$=
D$_{16} $=D$_{23}$=D$_{34}$=D$_{65}$=3.4meV.

\vspace{5mm}
{\bf Table 2.} {\small Average values of the spin projections of the
ions constructing a model six-spin cluster}

\begin{center}
\begin{tabular}{|c|c|c|c|c|c|c|}
\hline
$D_{ij}$ & $S_5$ & $S_6$ & $S_1$ & $S_2$ & $S_3$ & $S_4$ \\ \hline
0 & 0.16 & $-$0.118 & 0.614 & $-$0.145 & 0.007 & $-$0.017 \\
see text & 0.117 & $-$0.125 & 0.591 & $-$0.152 & 0.013 & $-$0.02 \\ \hline
\end{tabular}
\end{center}
\vspace{5mm}

Thus  the  results  obtained  on  the  basis  of  the   six-spin-model
correspond to the observed gapless ESR spectrum with the  anisotropic
$g$-factor   deviating   from   the   free   spin   $g$-factor.   The
five-spin-model does not  correspond to  the observed  ESR signals due
to the  absence of  a gap  in the  observed spectrum.   Therefore the
spectra  obtained  confirm  the  proposed  $S=$1/2  structure  of the
cluster.

Note that in our  experiments we observed the reduction of $g$-factor
with  temperature  for all  principal  orientations  of the  magnetic
field, while the model with collinear vectors ${\bf D}_{ij}$ predicts
the deviation for only one principal  direction, the  $g$-factors for
the  other   two   principal   directions   should  be  close  to  2.
Nevertheless  these simple models  demonstrate the possible
mechanism for the  reduction  of the value of effective  $g$-factor
and of its strong anisotropy.  The deviation from the free spin
$g$-factor value of the other  components  of the  $g$-factors
tensor may be probably provided by any noncollinearity of the
vectors ${\bf D}_{ij}$.

The nonlinearity  of magnetization  curves at  low temperatures which
is shown in  Figure~10 also confirms  the formation of  spin clusters
with their intrinsic  degrees of freedom,  because the energy  levels
of cluster are separated by the gaps depending in a nonlinear way  on
magnetic  field.   The  existence  of  a nonlinear susceptibility was
reported  as  evidence   for  Dzyaloshinski-Moriya  interactions   in
three-ion clusters in~\cite{Bazhan}.  The growth with temperature  of
the  linear  part  of  the  susceptibility,  visible in Figure~10, is
obviously due to the  temperature dependence of the  concentration of
the triplet  excitations of  the spin-Peierls  state.   The nonlinear
contribution to the magnetization  curves diminishes and vanishes  at
the temperature rise  from 1.6 to  6~K. At the  temperature about 9~K
the  magnetization  curve  becomes  again  nonlinear  but  with   the
convexity  directed  down.   The  differential  susceptibility grows
with  magnetic  field.   This  behaviour  may  be  explained  by  the
destruction   of   the   spin-Peierls   state   by  applied  magnetic
field~\cite{HaseMagnField}.   This destructive  magnetic field  could
be of the moderate value when the temperature is close to T$_{SP}$.

   The nonlinear magnetization curve at 2~K might be described as a
   sum of the linear $M(H)$ curve and of the magnetization of a
   paramagnet with the concentration of 10$^{-3}$ $S=$1/2 ions per
   one Cu-ion, the nonlinearity  being ascribed to the paramagnetic
   saturation. However the analogous curve for 6~K could not be
   described in this way because the  characteristic field of the
   saturation is not shifted to higher fields.

The change of the $g$-factor of Ni-doped samples with temperature may
be ascribed  to the  freezing  out of the  reservoir  of the  triplet
excitations of the spin-Peierls  state below T$_{SP}$.  It results in
the  switching-off  of the exchange  narrowing and in the breaking of
the collective character of the precession mode of the impurities and
triplet  excitations  of the  spin-Peierls  state.  This  process  is
similar to the temperature evolution of the resonance spectrum of the
triplet  excitations  in ion  radical  salts  with a  singlet  ground
state~\cite{Chesnut,McConnell}.  At   low   temperatures   only   the
impurity  mode  (spin-cluster  mode)  survives,  which  leads  to the
unusual  value of the  g-factor.  At  intermediate  temperatures  the
intermediate  $g$-factor is observable, due to the exchange mixing of
the spins states.

\section{CONCLUSION	}

The data obtained correspond to the x-T phase diagram show the linear
dependence of T$_{SP}$ and T$_N$ with doping concentration.  From the
AFMR   spectra   we  found  that  the   impurity   induced
antiferromagnetism   in   Cu$_{0.968}$Ni$_{0.032}$GeO$_3$    can   be
described  in terms of a  molecular  field  theory as a  conventional
orthorhombic  antiferromagnet  with the easy axis directed along $a$,
hard axis  along  $b$ and  second-easy  axis  along  $c$.  From  AFMR
spectra  we  obtained  gaps of 22 and  33~GHz and  anisotropy  fields
($H_{a1}\sim$ 0.017~T, $H_{a2}\sim$0.036~T).The antiferromagnetic
ordering of the  1.7\%~Ni-doped  sample  could not be  considered
within  the molecular  field  approximation  because  of the  small
value of the sublattice  magnetization  accompanied by the magnetically
disordered spin-Peierls  background.  This  fact  results  in the
absence  of a conventional  AFMR  spectrum.  The  anomalous  value and
temperature dependence  of  $g$-factor  in the  spin-Peierls  state  of
Ni-doped crystals  indicates the formation of spin clusters  around the
doping magnetic  ions.  The  presence of  magnetic  clusters  with
their own internal degrees of freedoms is confirmed by observation of
nonlinear magnetization  curves in the magnetically  disordered state.
Further theoretical and experimental  investigations of the cluster
structure are necessary for a more detailed  interpretation  of ESR
spectra and magnetization curves.

\section{ACKNOWLEDGMENTS}

The   authors   are   indebted   to   L.~A.~Prozorova,   S.~S.~Sosin,
I.~A.~Zaliznyak,   M.~A.~Teplov   and   M.~V.~Eremin   for   valuable
discussions.

This work is  performed  under the support of the Russian  foundation
for  fundamental  research  (grant number  98-02-16572)  and Civilian
Research and  Development  Foundation,  award N RP1-207.  Work at the
University   of  Warwick  was  supported  by  the  EPSRC  grant  {\it
Correlated Magnetic Systems} GR/K54021.

\section{Appendix}

The calculations of the energy levels of five ion linear cluster
Cu$_5$-Cu$%
_4 $-Ni$_1$-Cu$_2$-Cu$_3$ are based on the Hamiltonian \ref{Ham} The
energies of three low laying triplets with S=1  are
given approximately by \\

\begin{equation}
E_{1,2}=\frac{I_1+I_{23}}2+\frac{I_2+I_3}4\mp \frac 12\sqrt{(I_1-
I_{23}-%
\frac{I_2+I_3}2)^2+2(I_{12}+I_{13})^2}.
\end{equation}

\begin{equation}
E_3=\frac 12(I_3+I_2)-I_{23}.
\end{equation}
where\\
\begin{equation}
I_1=-\frac 34(J_{23}+J_{45}).
\end{equation}

\begin{equation}
I_2=-\frac 12(J_{14}+J_{15})-\frac 34J_{23}+\frac 14J_{45}.
\end{equation}

\begin{equation}
I_3=-\frac 12(J_{12}+J_{13})-\frac 34J_{45}+\frac 14J_{23}
\end{equation}

\begin{equation}
I_{12}=\frac 1{\sqrt{2}}(J_{14}-J_{15}),I_{13}=\frac 1{\sqrt{2}%
}(J_{12}-J_{13}),I_{23}=\frac 14J_{24}.
\end{equation}
The Dzyaloshinski-Moriya interaction splits the ground state
triplet  into the levels with the energies $E_s$, $E_{a,b}$\\

\begin{equation}
E_s=E_{1},~~~
E_{a,b}=E_1-C_1^2\frac{({\bf G}_{12}-{\bf G}_{14})^2}{E_3-E_1}.
\end{equation}
where\\

\begin{equation}
{\bf G}_{12}=\frac 14({\bf D}_{12}+{\bf D}_{23}-{\bf D}_{13}),~~
{\bf G}_{14}=\frac 14({\bf D}_{14}+{\bf D}_{45}-{\bf D}_{15}).
\end{equation}

\begin{equation}
C_1=\frac{0.5(I_2+I_3)+I_{23}-E_1}{%
\{[0.5(I_2+I_3)+I_{23}-E_1]^2+0.5(I_{12}+I_{13})^2\}^{1/2}}.
\end{equation}

The energies in the magnetic field are determined by the secular equation\\

\begin{equation}
\left|
\begin{tabular}{lll}
$E_s-\varepsilon $ & $g_{eff}^x\beta H_x$ & $g_{eff}^z\beta H_z$ \\
$g_{eff}^x\beta H_x$ & $E_a-\varepsilon $ & $ig_{eff}^y\beta H_y$ \\
$g_{eff}^z\beta H_z$ & $-ig_{eff}^y\beta H_{y,}$ & $E_b-\varepsilon $%
\end{tabular}
\right| =0
\end{equation}

Here $g_{eff}^x$ , $g_{eff}^y$ ,and $g_{eff}^z$ are given by\\

\begin{equation}
g_{eff}^x=g_{eff}^z=[C_1^2g_{Ni}+\frac 12C_2^2(g_{Ni}+g_{Cu})]\left(
1-\frac
12\left( C_1\frac{{\bf G}_{12}-{\bf G}_{14}}{E_3-E_1}\right) ^2\right)
\end{equation}

\begin{equation}
g_{eff}^y=C_1^2g_{Ni}+\frac 12C_2^2(g_{Ni}+g_{Cu})-\left( C_1\frac{%
{\bf G}_{12}-{\bf G}_{14}}{E_3-E_1}\right) ^2\left(
C_1^2g_{Ni}+(0.5-C_1^2)(g_{Ni}+g_{Cu})\right)
\end{equation}
where

\begin{equation}
C_2=\frac{-(I_{12}+I_{13})/\sqrt{2}}{%
\{[0.5(I_2+I_3)+I_{23}-E_1]^2+0.5(I_{12}+I_{13})^2\}^{1/2}}.
\end{equation}

The Dzyaloshinsky-Moriya vector ${\bf G}_{12}-{\bf G}_{14}$ is
supposed to be parallel to the y-axis. When the external magnetic field
is perpendicular to the vector ${\bf G}_{12}-{\bf G}_{14}$ and
$E_s-E_a>>g\beta H$ the frequency-field dependence is quadratic.

\newpage

{\bf Figure captions}

Figure~1.  Temperature evolution of the ESR line for the 0.5\%
Ni-doped sample.  $H\parallel c$, $f$=36.7~GHz.  The arrow marks the
resonance field of the free electron spin ($g$=2).

Figure~2.  Temperature evolution of the ESR line for the 1.7\%
Ni-doped sample.  $H\parallel b$, $f$=36.0~GHz.  The arrow marks the
resonance field for $g$=2.  Narrow line is DPPH-mark.

Figure~3.  Temperature evolution of the ESR line for the 3.2\%
Ni-doped sample.  $H\parallel a$ (easy axis direction).  The arrow
marks the resonance field for $g$=2.

Figure~4.  Temperature dependence of the ESR linewidth for the 1.7\%
Ni-doped sample, $f$=36.4~GHz.

Figure~5.  Temperature dependencies of resonance fields for the 3.2\%
Ni-doped sample.  $f\approx$36~GHz.  The arrows mark the resonance
fields for $g$=2.

Figure~6.  The ESR spectrum of the crystal doped with 1.7\%-Ni $H
\parallel c$, $T$=1.8~K

Figure~7.  Temperature dependencies of $g$-factors.

Figure~8.  Spectrum of the antiferromagnetic resonance of
Cu$_{0.968}$Ni$_{0.032}$GeO$_3$ at $T$=1.8~K.

Figure~9-a,b.  Temperature dependencies of the integrated ESR
intensity for pure and doped crystals. Dotted curve is the
the susceptibility of a paramagnet containing
0.5\% of  spins S=1, $g=$2.3.

Figure~10.  Magnetization curves for the 1.7\% Ni-doped sample at
$H\parallel a$.

Figure~11.  The structure of the dimerised lattice around the
impurity atom.  The five- and six-spin-clusters. The arrows indicate
the direction and the value of the average spin projections within
the six-spin-cluster with $D_{ij}=$0.

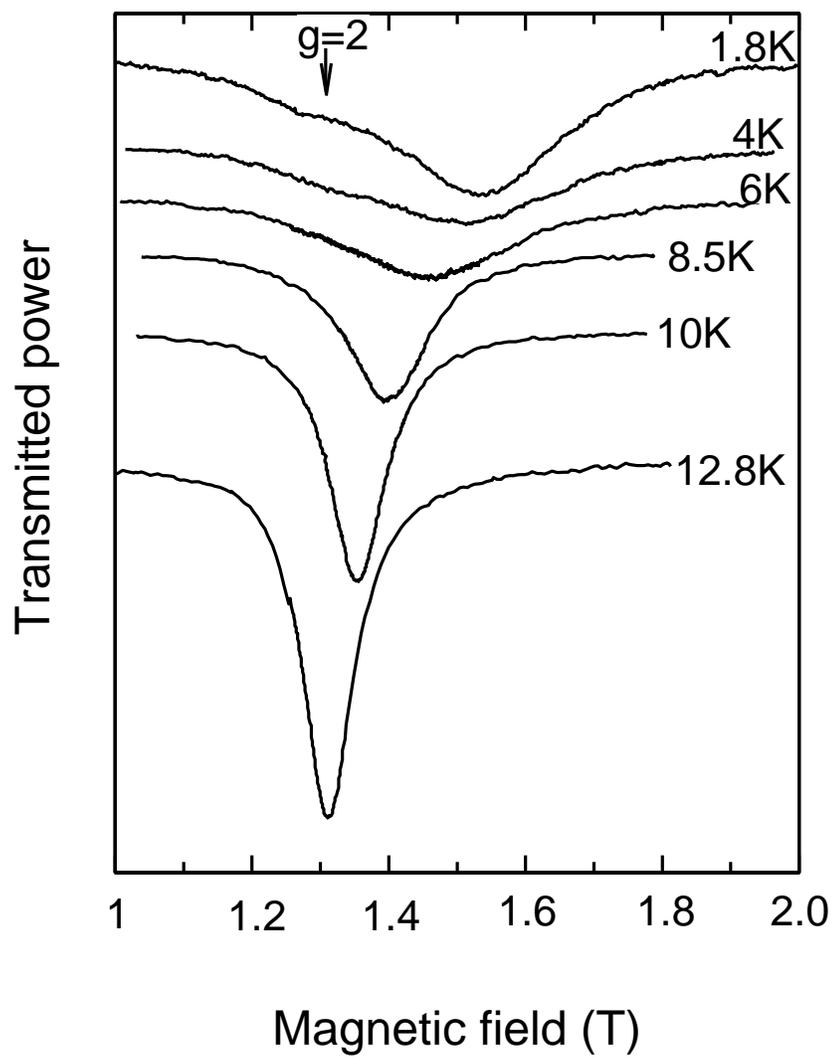
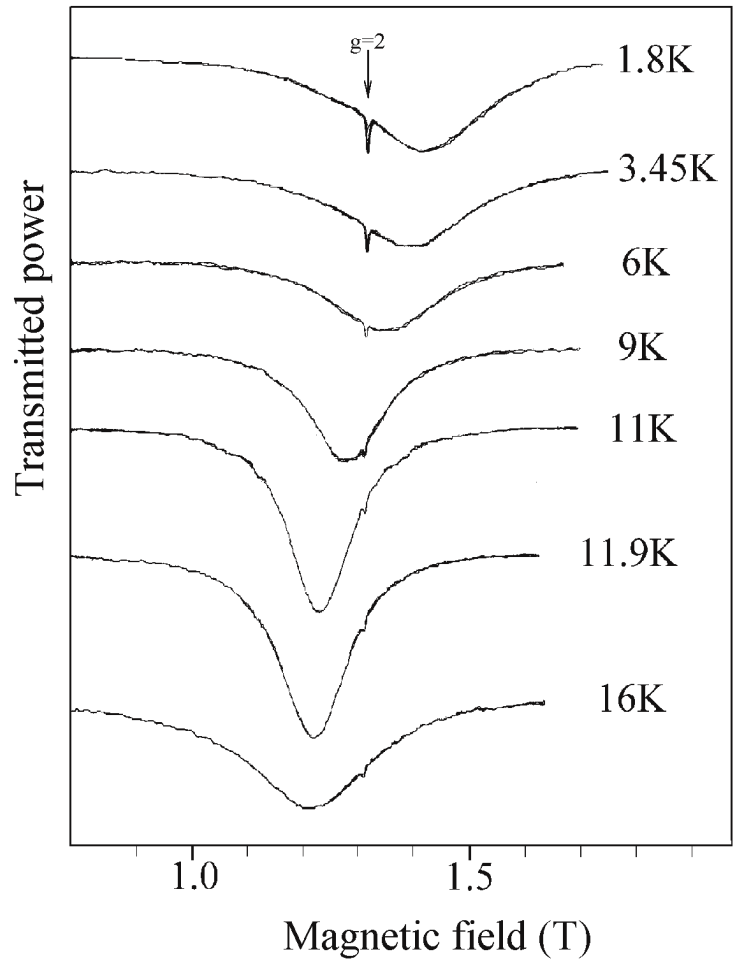
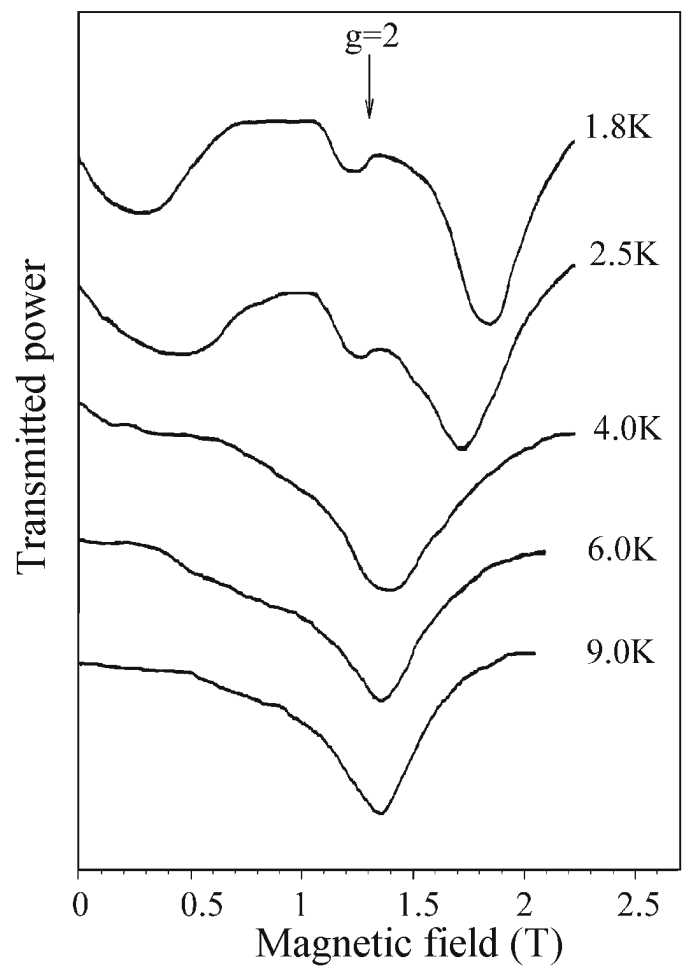
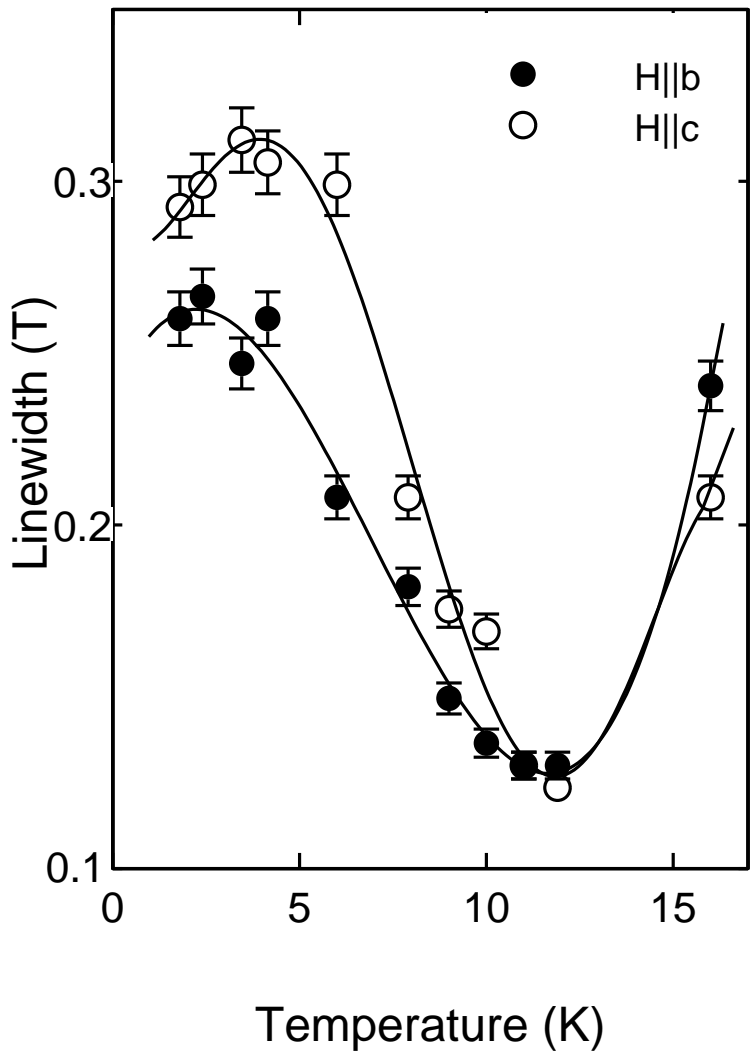
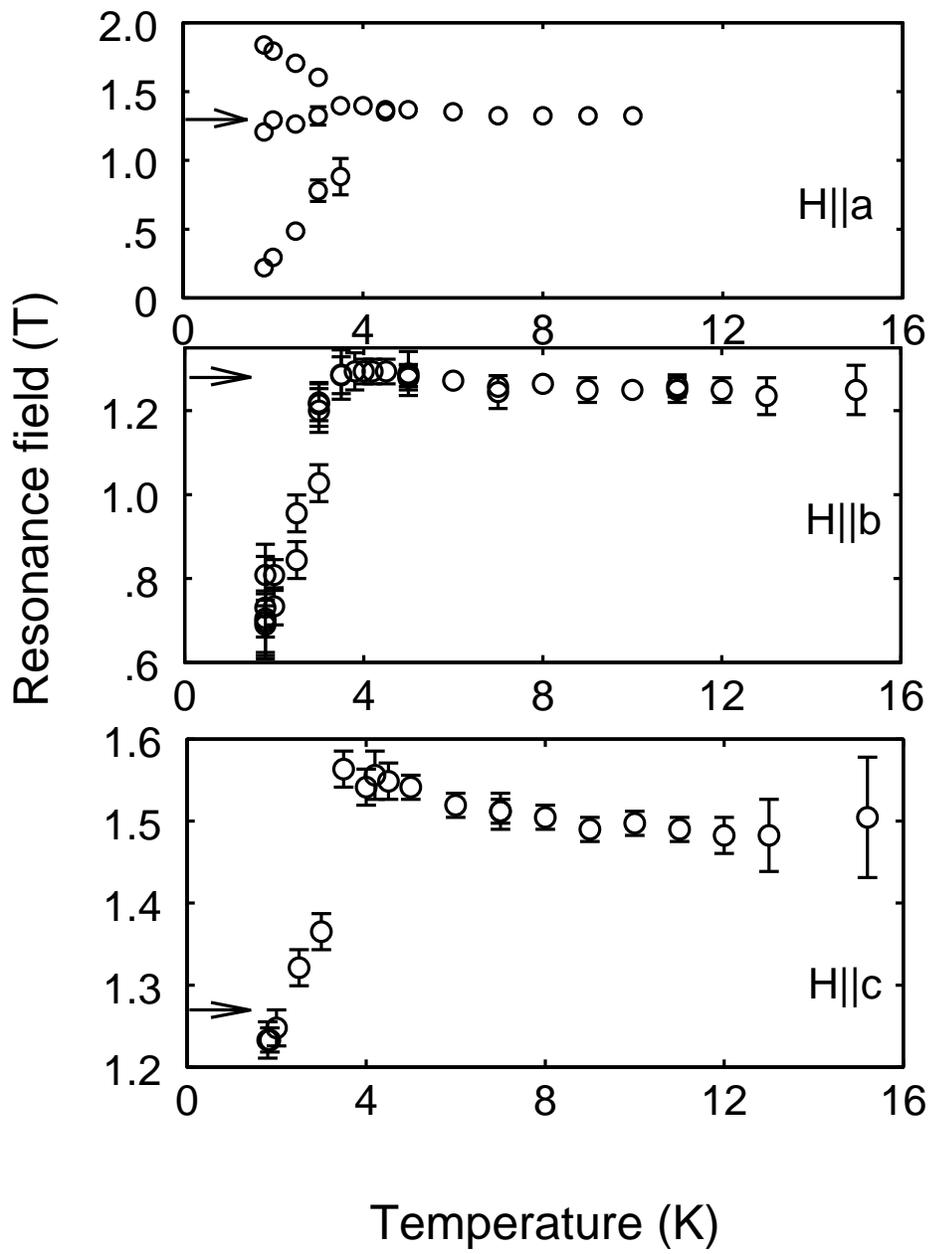
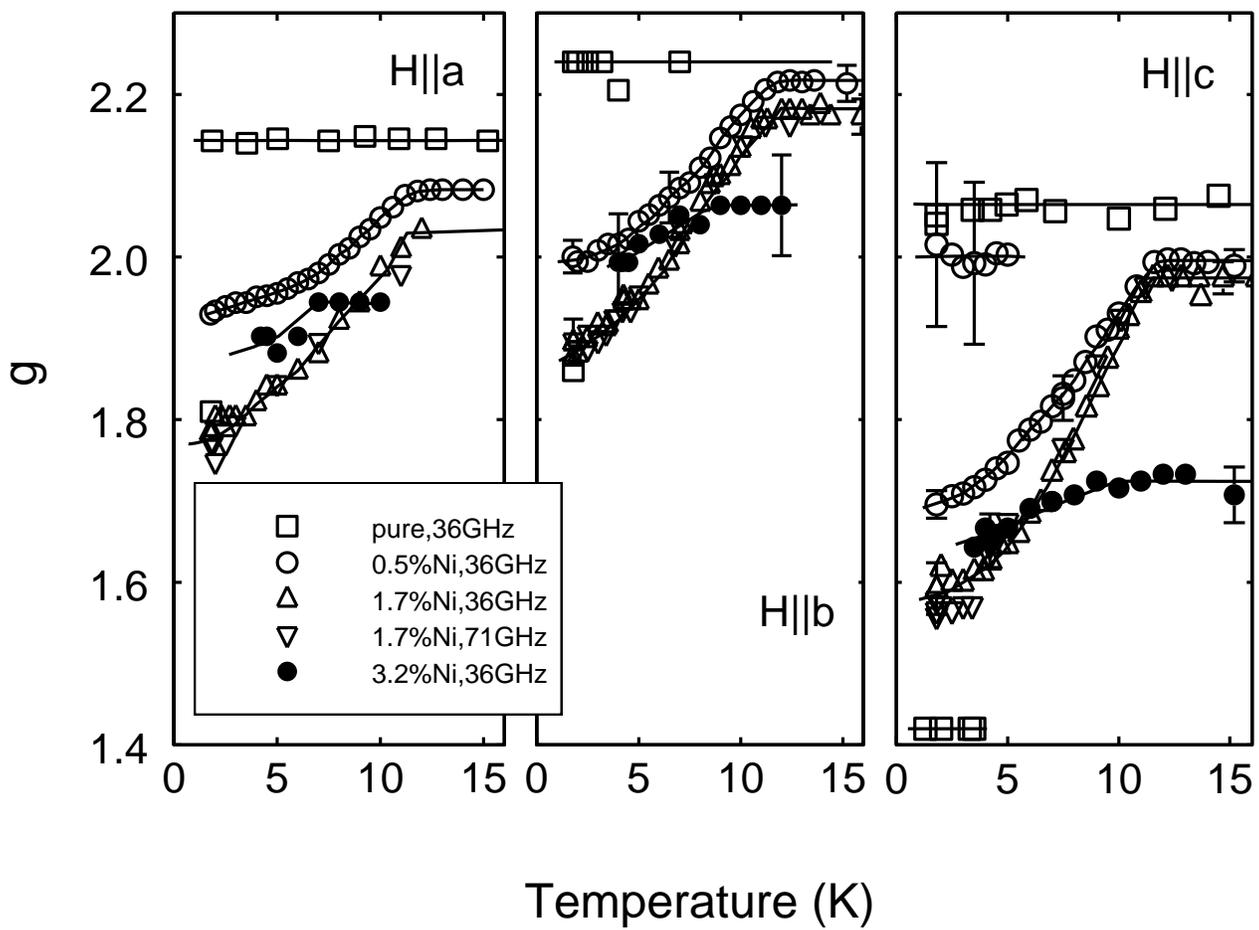
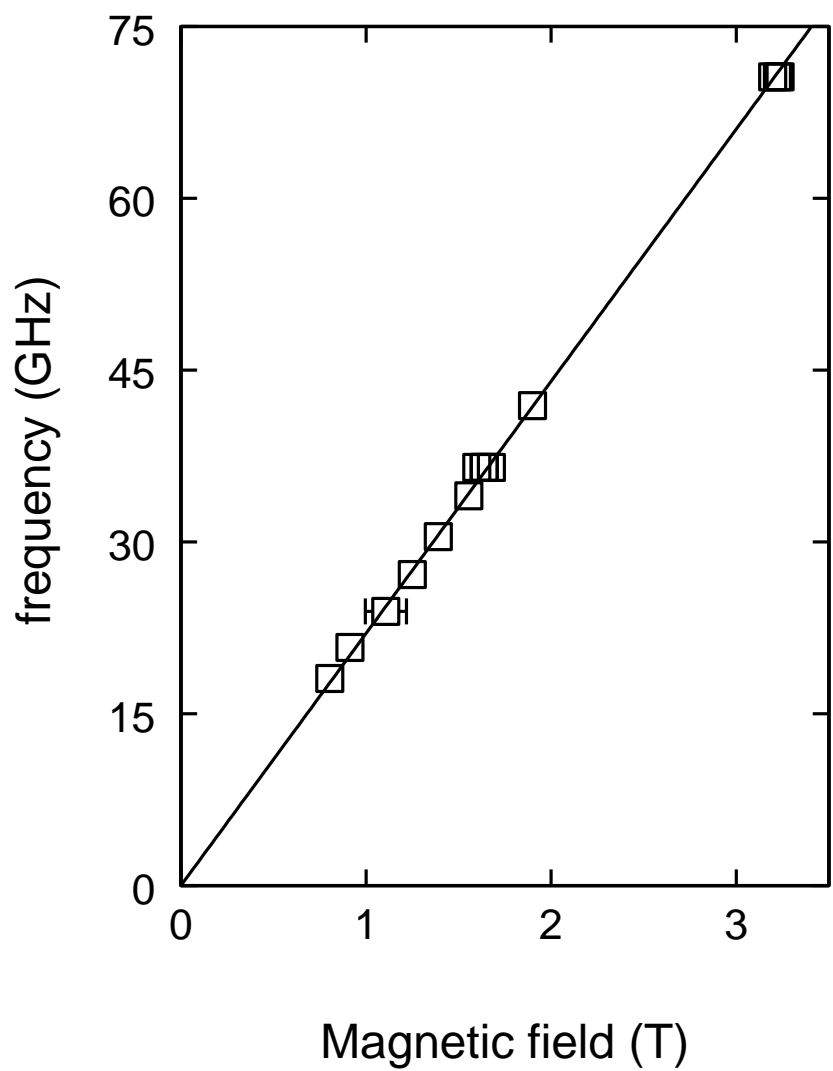
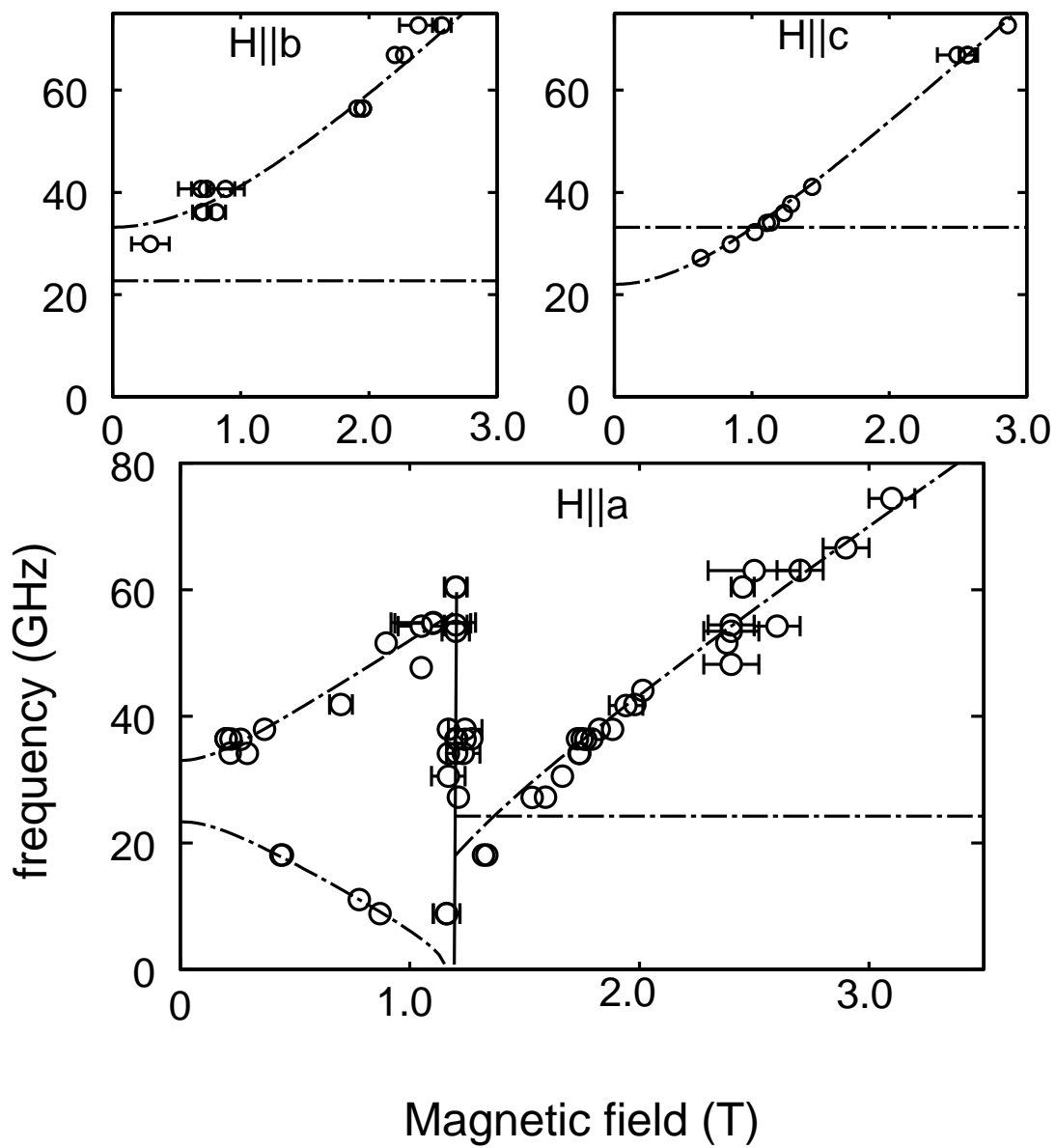
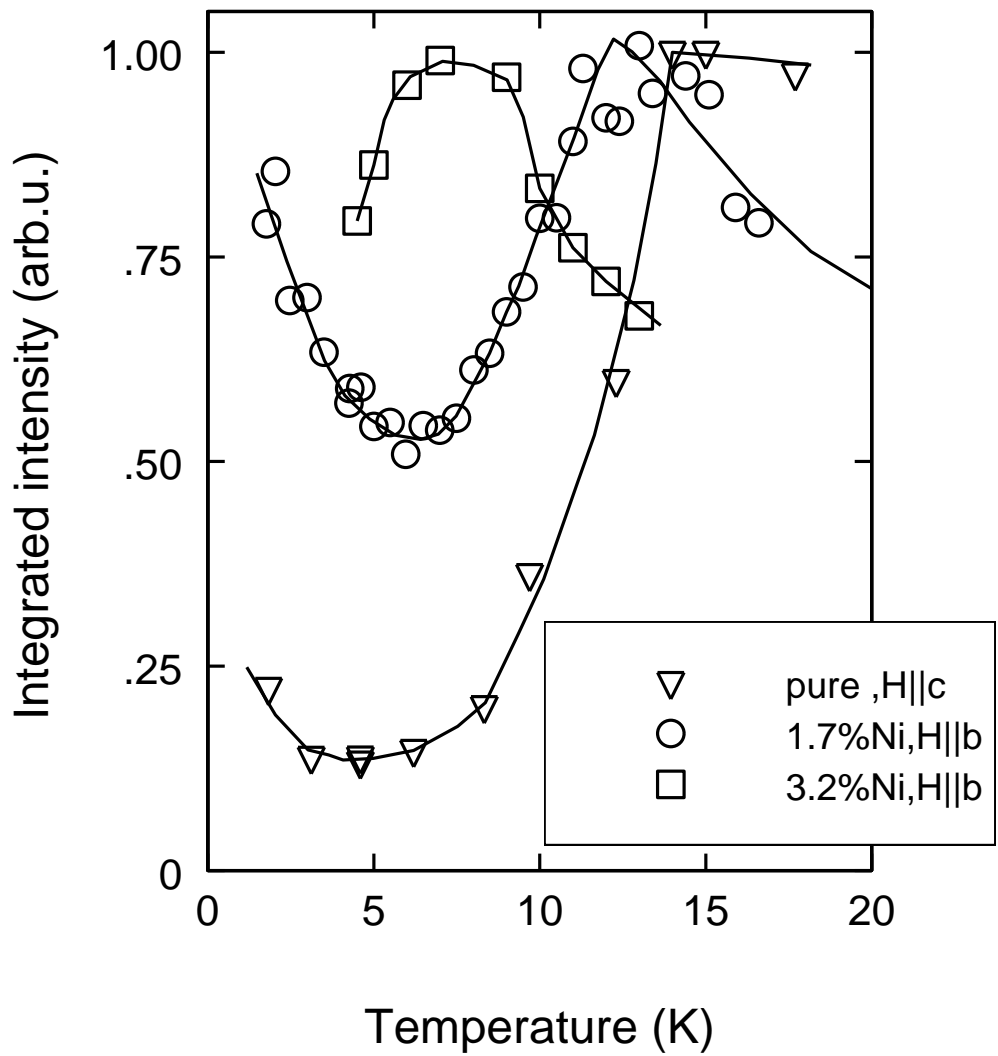
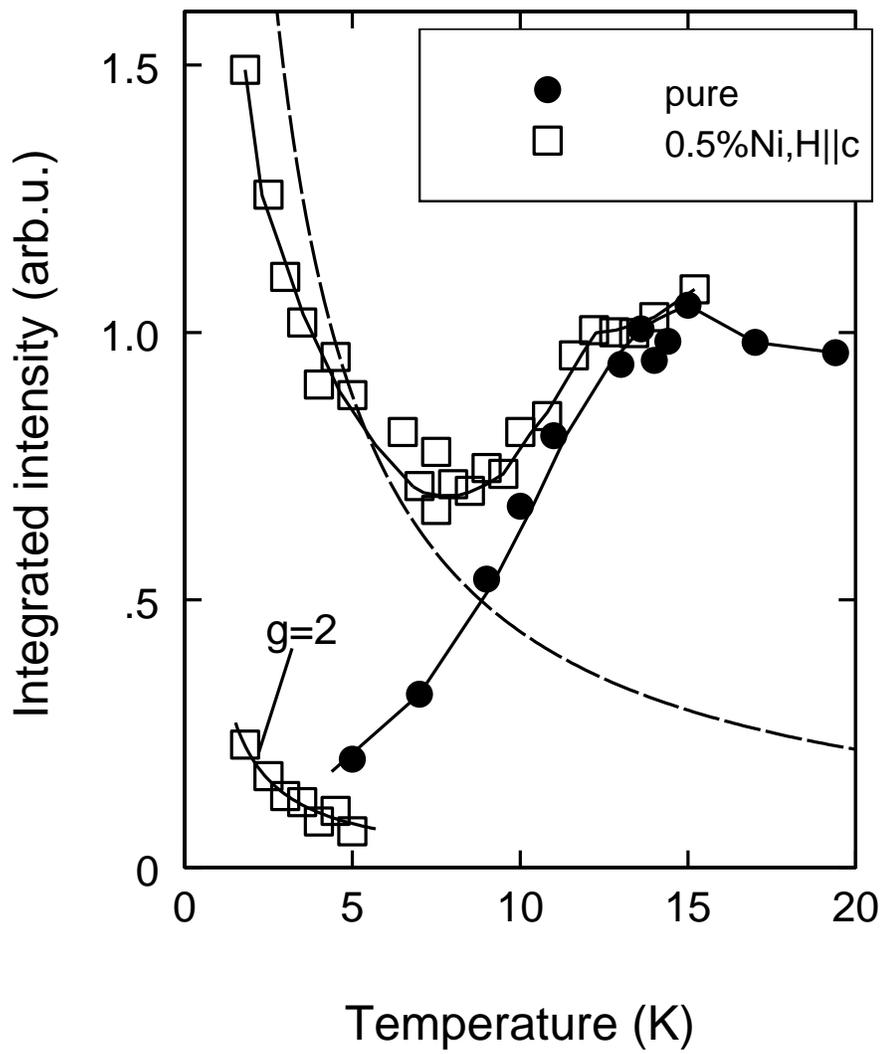
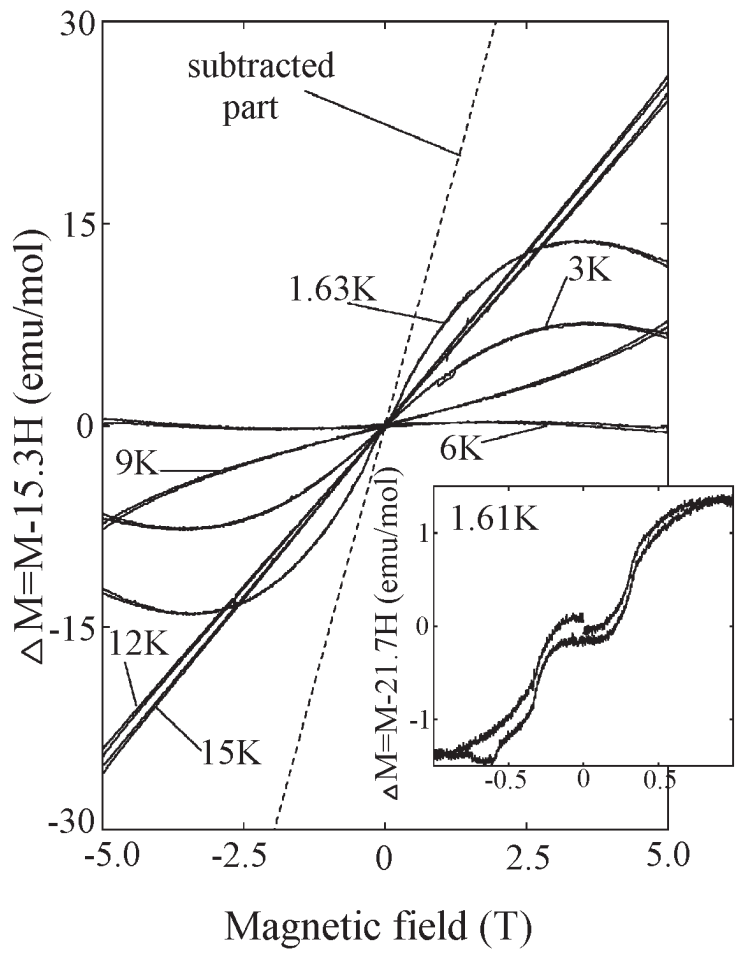
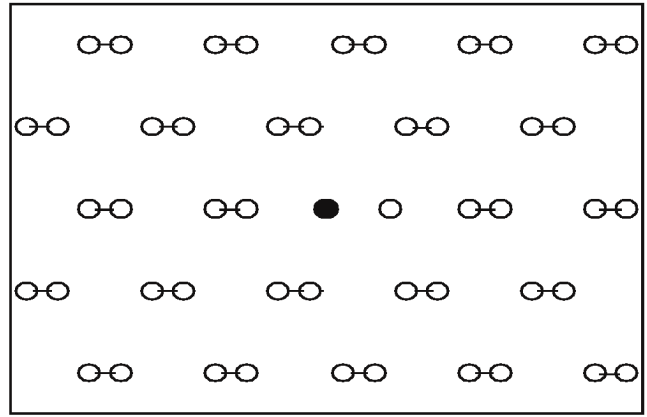
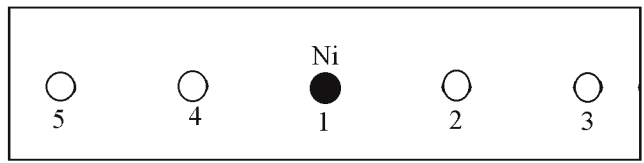
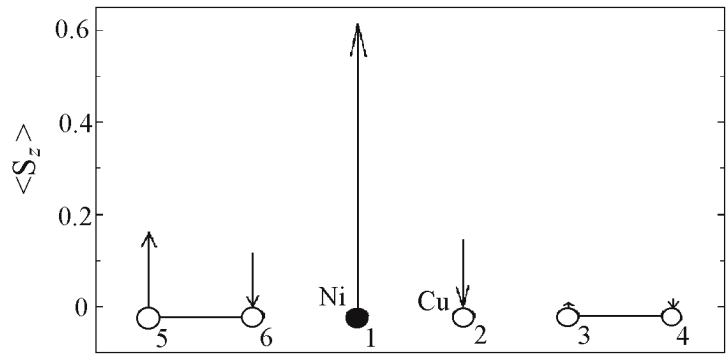
\end{document}